# How Reliable and Useful Is Cabell's Blacklist? A Data-Driven Analysis

**Christophe Dony**

ULiège Library, University of Liège, Belgium
cdony@uliege.be; https://orcid.org/0000-0001-7421-5993

**Maurane Raskinet**

ULiège Library, University of Liège, Belgium
maurane.raskinet@uliege.be; https://orcid.org/0000-0002-0394-0167

**François Renaville**

ULiège Library, University of Liège, Belgium
francois.renaville@uliege.be; https://orcid.org/0000-0003-1453-1040

**Stéphanie Simon**

ULiège Library, University of Liège, Belgium
stephanie.simon@uliege.be; https://orcid.org/0000-0002-9591-9946

**Paul Thirion**

ULiège Library, University of Liège, Belgium
paul.thirion@uliege.be; https://orcid.org/0000-0002-2736-4707

## Abstract

In scholarly publishing, blacklists aim to register fraudulent or deceptive journals and publishers, also known as "predatory", to minimise the spread of unreliable research and the growing of fake publishing outlets. However, blacklisting remains a very controversial activity for several reasons: there is no consensus regarding the criteria used to determine fraudulent journals, the criteria used may not always be transparent or relevant, and blacklists are rarely updated regularly. Cabell's paywalled blacklist service attempts to overcome some of these issues in reviewing fraudulent journals on the basis of transparent criteria and in providing allegedly up-to-date information at the journal entry level. We tested Cabell's blacklist to analyse whether or not it could be adopted as a reliable tool by stakeholders in scholarly communication, including our own academic library. To do so, we used a copy of Walt Crawford's Gray Open Access dataset (2012-2016) to assess the coverage of Cabell's blacklist and get insights on their methodology. Out of the 10,123 journals that we tested, 4,681 are included in Cabell's blacklist. Out of this number of journals included in the blacklist, 3,229 are empty journals, i.e. journals in which no single article has ever been published. Other collected data points to questionable weighing and reviewing methods and shows a lack of rigour in how Cabell applies its own procedures: some journals are blacklisted on the basis of 1 to 3 criteria – some of which are very questionable, identical criteria are recorded multiple times in individual journal entries, discrepancies exist between reviewing dates and the criteria version used and recorded by Cabell, reviewing dates are missing, and we observed two journals blacklisted twice with a different number of violations. Based on these observations, we conclude with recommendations and suggestions that could help improve Cabell's blacklist service.

## 1. Introduction

As academic librarians, we are often confronted with questions from researchers that are unsure about the quality or serious character of particular open access journals. We therefore help and train our research community to critically examine a journal's publishing practices and policies as they relate to, for example, peer review, editorial services, indexing services, metrics, open access policies, and the type of article processing charges (APCs) required by the publisher, if any. We also remind researchers of the existence of whitelists (e.g., the Directory of Open Access Journals – DOAJ) and of the campaign *Think*, *Check, and Submit*. Most of the time, this type of work is useful in identifying whether or not a journal is fake or deceptive, that is whether or not it can be considered 'predatory'[1] because it requires the payment of fees while deliberately "sow[ing] confusion" and clearly deceiving both readers and authors in "deviat[ing] from best editorial and publication practices" (Grudniewicz et al., 2019).

In some cases, however, the type of reviewing work described above can be a hard and tricky task. A recent cross-sectional analysis (Strinzel, Severin, Milzow, & Egger, 2019) of how whitelists (e.g. DOAJ) and blacklists[2] (e.g. Stop Predatory Journals) may help the scholarly community to tackle fraudulent publishing, for instance, shows that some journals and publishers appear to be in a gray area as they are included in both whitelists and blacklists at the same time. Moreover, some fake journals have managed to integrate some tools and services of third-party providers such as bibliographic databases, to which most academic libraries subscribe (see Manca, Cugusi, Dvir, & Deriu, 2017; Nelson & Huffman, 2015; Somoza-Fernández, Rodriguez-Gairón, & Urbano, 2016).

This may be even more puzzling for academic librarians and researchers when third-party services provide rather vague explanations as to how they tackle the phenomenon of deceptive publishing. In their collection policies, for example, Ulrich's Periodicals Directory (ProQuest)[3] state that the "identification and screening of [...] 'vanity' titles is an ongoing editorial effort that can be complicated by the fact that some have obtained abstracting and indexing coverage in commercial services" (ExLibris, 2019a). Although they specify that they continue "to monitor these types of titles" and that they work "together with the library community to identify specific publications about which libraries may have concerns" (ExLibris, 2019a), it is not clear if they aim to exclude fraudulent journals from their directory, nor if they work to remove fake journals that may have already found a way in said directory. Clearer is ExLibris' statement that announces that they "will remove any content that is determined to be predatory from [their] products" (ExLibris, 2019b), namely products like the Alma Community Zone and the Central Discovery Index, exactly like EBSCO (EBSCO, 2019). How they will do so, however, remains unknown.

Third-party service providers' lack of transparency regarding how they aim to combat fake publishing outlets is not surprising; it reflects a larger struggle that other stakeholders in scholarly publishing face when dealing with deceptive journals and publishers: there is no widely approved definition for deceptive journals and publishers (Grudniewicz et al., 2019). As a result, checklists aiming to identify fake publishing outlets abound and no single checklist is widely approved. Moreover, the few checklists that are used to compile blacklists do not always distinguish poor-quality journals from deceptive ones. Finally, blacklists are not regularly updated. In the meantime, however, fake publishing outlets keep undermining sound and rigorous science and continue to harm academic authors and readers, the general public, funders, learned societies, academic libraries, as well as the Open Access movement (see Eve & Priego, 2017). This is why a sound, common, and reliable framework for identifying and combating fake journals and publishers that manages to overcome some of the issues mentioned above remains highly necessary.

This is pretty much the promise behind Cabell's blacklist service. Since June 2017, Cabell's paywalled blacklist service has used over 60 transparent criteria "to evaluate all journals suspected of deceptive, fraudulent, and/or predatory practices" (Toutloff, 2019b). The blacklist provides allegedly up-to-date information at the journal entry level for all of the offenses witnessed after review, which should constitute a useful added-value compared to other existing blacklists. Cabell is not new in the field of scholarly publishing; it is a for-profit American company that has helped researchers and universities to evaluate and examine academic journals by maintaining a whitelist for almost 40 years. We therefore thought that the tool was worth considering to help us and our research community to identify fake and deceptive journals more rapidly and effectively.

Before considering a subscription to Cabell's blacklist, we decided to run a test of their service to see whether it lived up to its promises. Cabell granted us a two-week trial period, from August 26, 2019 to September 8, 2019. During that period, we tested 10,123 journals from a copy of Walt Crawford's Gray OA 2012-2016 dataset (Crawford, 2016), which consists of a detailed inventory of potentially 'predatory' journals included in Jeffrey Beall's lists as they appeared on his blog as of July 8, 2016 – Beall is the one who coined the term 'predatory publishing' in 2010 (Beall, 2010).

This sample test and the data that we collected for journals included in Cabell's blacklist has provided us with a better understanding of the company's service, its coverage, and methodology, including minimal requirement for inclusion in the blacklist. This article details how we performed this analysis and discusses major flaws found in Cabell's blacklist, particularly as they relate to weighing method and minimal requirements for inclusion, journal prioritisation, and the soundness of Cabell's own procedures.

## 2. Literature Review

Jeffrey Beall has been a central figure in the debates surrounding the phenomenon of deceptive publishing in academia. For almost a decade, the American librarian compiled and maintained lists of so-called "potential, possible, or probable predatory" open access journals and publishers on his blog. By the end of 2016, Beall's blacklists of stand-alone journals and publishers contained more than 1,000 entries each. In early 2017, however, Beall shut down his blog where the lists appeared for reasons that are still unclear.[4]

What is clear is that Beall's lists were not flawless, and his work far from unbiased. First, Beall has always been adamantly critical of and reactionary towards the Open Access movement (Beall, 2013, 2017). His perceived image of the world of scholarly publishing has been characterised as skewed and conservative (Olivarez, Bales, Sare, & vanDuinkerken, 2018; Swauger, 2017). And his work has been said to discriminate against non-Western researchers and publishers (Berger & Cirasella, 2015). Second, "false positives" have made their way into Beall's blacklists. In 2013, scientific journalist John Bohannon reported the results of his "sting" operation on Open Access publishers (Bohannon, 2013), which consisted in him sending fabricated fake scientific articles to 304 Open Access journals requiring APCs (article processing charges), most of which were included in Beall's lists. A vast majority of the journals indexed in Beall's lists (82%) accepted the fake paper(s). But a bit less than 20% of these journals did not, which showed that Beall's lists were not foolproof. The sheer volume of deceptive publishing outlets suggested by Beall's lists has also seriously been challenged. Crawford has shown that out of the 18,910 journals entailed in Beall's lists, 10,019 were empty shells with no articles ever published, and therefore argued that Beall overtly exaggerated the quantitative significance of the phenomenon of deceptive publishing (Crawford, 2017). Finally, Beall's set of criteria for blacklisting journals has been severely called into question for being subjective and not allowing to distinguish between truly deceptive journals and poor-quality ones (Olivarez et al., 2018). This is also why other checklists and approaches to identify "predatory" journals have surfaced (see Cukier et al., 2020; Eriksson & Helgesson, 2017; Laine & Winker, 2017; Schmitz, 2019).

Despite its severe limitations, Beall's work has nevertheless proved useful for two reasons. First, his work has raised global awareness of the phenomenon of deceptive publishing in academia. Second, his lists have been used as cross-verification tools for the contents of other so-called whitelists, including the Directory of Open Access Journals (DOAJ). In 2014, Walt Crawford compared the contents of Beall's lists with that of the DOAJ and found that this latter directory then contained some 900 titles that were indexed in Beall's lists (Crawford, 2014). This represented almost 10% of the 2014 contents of the DOAJ. While this overlap did not specifically mean that the DOAJ indexed so-called "predatory" journals, it nevertheless raised serious concerns regarding DOAJ's quality-control verifications. These concerns were also heightened by Bohannon's previously cited sting operation, which showed that 73 journals that had accepted fake fabricated papers were included in the DOAJ (Bohannon, 2013). Most likely as a result of this controversy, the DOAJ strengthened its application procedure and standards for inclusion in 2014, which eventually led to a directory-wide re-application process (Van Noorden, 2014). This re-application process ran from January 2015 to December 2017 (DOAJ, 2017).

Both blacklists and whitelists certainly have their imperfections, just as they both have their detractors (Matumba et al., 2019; Teixeira da Silva & Tsigaris, 2018). But it is difficult to argue that both are not necessary. They serve different objectives, which should not be considered mutually exclusive. Only when used co-jointly can whitelists and blacklists better help libraries, funders, researchers, and third-party services to navigate the sometimes muddy waters of scholarly publishing. Moreover, it is important that efforts be made to regularly improve and update blacklisting practices. Fraudulent publishing practices change over time and therefore need ongoing debates such as, for example, the discussion document on predatory publishing released by The Committee on Publication Ethics (COPE, 2019).

Cabell's blacklist seems to have understood the need for a more rigorous and ongoing approach to blacklisting and promises to offer a good alternative to Beall's lists for various reasons. Contrary to many websites that resurrected Beall's lists in one form or another after Beall shut down his blog, Cabell's blacklist provides a truly dynamic indexing service. Their staff continuously verify potentially deceptive journals according to a transparent set of criteria – to date Cabell has produced two transparent criteria versions: v1.0 and v1.1. Cabell's v1.0 is the first checklist compiled by the company to evaluate the potential deceptive nature of journals. It contains 64 criteria, which are organised by subject matter such as "integrity", "peer review", and "publication practices" (Toutloff, 2019a). This v1.0 evaluation list was used for the preparation of the blacklist before it was launched until early 2019 when Cabell resorted to a new v1.1 criteria version, which officially went into effect on March 13, 2019. This v1.1 evaluation checklist features 74 behavioural indicators, which are grouped "according to relative severity and subject matter" (Toutloff, 2019b).

Much more so than Beall, Cabell has also provided some information regarding their weighing method for blacklisting journals. In an industry update, Cabell's Communications Manager Michael Bisaccio explains that through a "careful analysis" of deceptive behaviours, "a rubric was created and applied in the investigation of each journal", which "produced a weighted score whose magnitude increased with the probability that a journal was engaging in deceptive behaviours" (Bisaccio, 2018, p. 246). Each confirmed deceptive marker is thus assigned points according to its gravity, and a "total score over 100 is the threshold for including a journal" in the blacklist (Bisaccio, 2018, p. 246). Moreover, in contrast to Beall who updated two lists – one for fake stand-alone journals and another one for fake publishers, Cabell organises its blacklist at the level of journal entry. For each journal entry, detail of violations, the date of the last review, and the criteria version used are indicated. Another welcome addition noted by academic librarian Rick Anderson is Cabell's inclusion of an appeals policy (Anderson, 2017), though this appeals policy seems difficult to find (Anderson, 2019) and is only available to subscribers.

Finally, Cabell's blacklist also tries to overcome another major flaw underlying Beall's lists and methodology, namely subjectivity and discrimination against non-Western publishing outlets. Bisaccio indeed notes that the scoring method of Cabell's blacklist "was designed specifically to ensure that legitimate journals that are new, from developing countries, or are simply low quality, are not classified as 'predatory' and included in the Blacklist" (Bisaccio, 2018, p. 246).

Overall, because Cabell has produced public criteria versions and has provided some information regarding their operationalisation, its blacklist "should be less susceptible to charges of subjectivity, arbitrariness and bias, which Beall faced with his blacklist" (Siler, 2020, p. 4).

Commentators have nevertheless pointed out issues with Cabell's blacklist. Technical problems and limitations with the blacklist's advanced search mode have been reported (Anderson, 2017, 2019; Callicott, 2015). Anderson has also expressed concerns regarding the lack of clarity of particular criteria, and how some of them remain difficult to verify and can be considered "inferential" (Anderson, 2019).

Next to these issues, studies still need to show the reliable character underlying the data behind Cabell's blacklist itself, especially since it seems to have started to function as a replacement capacity for Beall's lists in scholarly research focused on fake and deceptive publishing outlets (see Anderson, 2020; Severin, Strinzel, Egger, Domingo, & Barros, 2020; Siler, 2020). In this respect, electronic services librarian Xiaotian Chen has already noted that Cabell has surprisingly indexed journals that have now ceased to exist or became inactive, in comparing data between Beall's blacklists and that of Cabell (Chen, 2019). Our study aims to further question how useful and reliable is Cabell's blacklist. In doing so, we shed further light on Cabell's procedures and criteria operationalisation.

## 3. Methods

Our primary objectives for this study were to assess the coverage of Cabell's blacklist, to examine the types of journals indexed in it, and to better understand the methodology and procedures Cabell uses for blacklisting journals. To do so effectively and methodically, we used Walt Crawford's Gray OA 2012-2016 dataset (Crawford, 2016), which we used to proceed to a partial comparison with Cabell's blacklist.

Crawford's dataset is well suited for these objectives. First, it is – to date – the most comprehensive listing of potentially or probably fake individual journal entries; it contains 18,910 journal entries. Secondly, for each entry, Crawford provides useful journal metadata and article volume information (when retrievable): publisher, journal title, journal URL, APC amount, number of articles published yearly between 2012 and 2016[5], and journal's starting publication date. Finally, Crawford also assigns a specific code to each journal.[6] These codes provide insightful information regarding the status of journals and article volume (e.g. no articles published later than a specific year). These codes include "E" for empty, "XM" for malware, and "XX" for unreachable and can be used to distinguish various types or spectra of fraudulent journals according to specific characteristics.

To carry out our comparative analysis, we created a copy of Crawford's dataset, which we modified for our purposes. For practical reasons, including readability, we removed Crawford's original columns relating to APC amount, journal starting publication date, and the number of articles published yearly. But we retained Crawford's codes which, as mentioned above, provide useful information. We then added a column to identify whether or not the journal was included in Cabell's blacklist. We added three more columns to collect the following initial data as recorded by Cabell in case of inclusion in the blacklist: number of violations registered, the date when the journal was last reviewed, and the criteria version used – v1.0 or v1.1. Our dataset is available at the LIBER Quarterly Dataverse (part of the Harvard Dataverse), at https://doi.org/10.7910/DVN/IWJIRN. It can also be found at https://doi.org/10.5281/zenodo.4007501.

We then added extra columns to record other information after some initial data was collected and as we started to perform cross-verifications. We added an extra column 'New Code' in order to check and potentially change the 'Empty' code (E) that Crawford assigned to journals – by far Crawford's most used code in his dataset, which contains 10,009 journals tagged as empty (52.93% of the dataset). We used the new code 'NE' (Not Empty) in this column to retag journals that we observed were not empty anymore after verification. When this new code (NE) was assigned to a journal, the year in which the first article was published or the publication date of the journal's first volume was recorded in an additional column. We also added two other columns to better understand the types of violations used to blacklist journals indexed on the basis of a rather small or big number of violations. One column was aimed to record the detail of violations for two specific ranges of journals that formed the extremes of Cabell's blacklist violation spectrum as reflected in our sample, namely one segment of journals blacklisted on the basis of 1 to 3 violations, and another segment of journals blacklisted on the basis of 11 to 15 violations – 15 being the highest number of violations recorded in our sample. Because of time constraints, we were not able to record details of violations for the entirety of the second segment. Instead, we recorded details of violations for journals included in this second segment (11 to 15 violations) which contained repeated statements of identical violations, which we noticed during our cross-verifications. An extra column was thus added to verify whether or not journals blacklisted on the basis of 11 to 15 criteria contained identical violations recorded multiple times.

In order to check as many journals as possible within our trial period (from August 26$^{th}$, 2019 to September 8$^{th}$, 2019), we divided our copy of Crawford's dataset into segments of 500 lines and asked librarian colleagues[7] to help us verify initial data (see above). Out of the 18,910 entries from our dataset, 10,123 journals were tested.

A majority of the journals that were tested belong to the first 8,000 entries of our dataset, which follows Crawford's original dataset organisation, i.e. an alphabetically ordered classification by journal title. Quite logically, reviewers signed up for one or several sets of 500 journals following the dataset's order. For 420 of the journals reviewed within the first 8,000 lines of our dataset, data is absent. This is either because we realised after cross-verifications that data for some entries was erroneously collected, or because some reviewers did simply not have the time to finish reviewing the set(s) for which they had signed up within the allotted time. These 420 entries were therefore simply not included in our sample.

The rest of the journals that we tested (2,543) appear in various other sections of our dataset, i.e. beyond the first 8,000 lines, for different reasons. First, some journals were more or less randomly selected and tested simply because they belonged to a publisher that did not get any matches in Cabell's blacklist after its name was searched using the 'publisher' option in the advanced search function. In that case, we applied a filter in our dataset with said publisher to mark all of its journals as not included in Cabell's blacklist. This strategy was only resorted to after several journals published by the same publisher did not get any matches with a standard journal title search in Cabell's blacklist, so as to avoid possible discrepancies between the publishers' names used by Crawford and Cabell. Second, we were particularly interested in testing some specific titles for internal purposes, including some of the journals that were reported to be indexed in the DOAJ and in Cabell's blacklist (see Strinzel et al., 2019). Finally, another reason why some journals beyond the first 8,000 lines were selected for review is a bit more oriented and can be summarised in what follows. Over the course of our cross-verifications, we observed that many journals that were tagged as empty (E) by Crawford were included in Cabell's blacklist. We therefore decided to further select other so-tagged empty journals to assess their proportion in Cabell's blacklist. To facilitate this collection of extra data, we selected several publishers with a rather important portfolio whose majority of journals were tagged as empty and were not listed in the first 8,000 journals of our dataset.

Our sample selection method can be said to be biased because we did not apply a random sampling method. However, we believe that this bias is of limited significance. First, our selection of titles remains representative in terms of numbers of publishers tested. Our sample as a whole accounts for 64 % (398) of the total number of publishers appearing in Crawford's original dataset (622). Second, the proportions of journals as classified by Crawford's codes in our sample do not sharply differ from those found in Crawford's original dataset (see Table 1).

Our collection of data provides an image of Cabell's blacklist that is frozen in time. This can be considered a limitation in our assessment of Cabell's coverage. Cabell indeed keeps working continuously on the expansion of its blacklist. On August 29th, 2019, for example, the blacklist featured 11,839 titles. On October 2nd, 2019, however, Cabell reported on their blog that their blacklist had reached a peak of 12,000 journals (Linacre, 2019). More recently, on February 26th, 2020, Cabell's blog celebrated how the blacklist surpassed 13,000 journals (Linacre, 2020). It is therefore possible that the results presented below are now slightly inaccurate insofar as they do not reflect Cabell's dynamic indexing or the possible corrections that they may have made to some problematic journal entries. Nevertheless, we believe that this limitation is of little significance for two reasons. First, while the results regarding the number of journals included in Cabell's blacklist may be put into perspective, we still believe they provide fruitful insights for stakeholders in scholarly publishing. Second, most of the results below focus on data regarding titles that are included in Cabell's blacklist.

*Table 1: Comparative proportions of journals per code in Crawford's dataset and in our sample.*

| | Crawford's dataset | | Sample | |
|---|---|---|---|---|
| Journal Code (Crawford) | Number of journals per code | % of journals per code | Number of journals per code | % of journals per code |
| **A** (No other code assigned) | 3,795 | 20.07% | 1,250 | 12.35% |
| **B3** (No articles later than 2013) | 346 | 1.83% | 146 | 1.44% |
| **B4** (No articles later than 2014) | 552 | 2.92% | 218 | 2.15% |
| **B5** (No articles later than 2015) | 1,199 | 6.34% | 466 | 4.60% |
| **BC** (No articles later than 2012, or explicitly cancelled or merged) | 104 | 0.55% | 28 | 0.28% |
| **BF** (Fewer than three 2016 articles) | 820 | 4.34% | 316 | 3.12% |
| **BR** (Primarily conference proceedings) | 24 | 0.13% | 6 | 0.06% |
| **E** (Empty) | 10,019 | 52,98% | 6,999 | 69.13% |
| **UA** (Unknown or hidden APC) | 902 | 4.77% | 327 | 3.23% |
| **XH** (Hybrid) | 113 | 0.60% | 25 | 0.25% |
| **XM** (Malware) | 60 | 0.32% | 20 | 0.20% |
| **XN** (Not OA) | 135 | 0.71% | 33 | 0.33% |
| **XO** (Opaque; too difficult to count) | 72 | 0.38% | 13 | 0.13% |
| **XU** (Unworkable) | 23 | 0.12% | 11 | 0.11% |
| **XX** (Unreachable) | 746 | 3.95% | 265 | 2.62% |
| **Total** | 18,910 | 100% | 10,123 | 100% |

## 4. Results

Out of the 10,123 journals that we tested from Crawford's dataset, 4,681 are included in Cabell's blacklist. This represents 46.24% of our sample. At the time of this study, this number also accounts for 39.01% of Cabell's blacklist as based on their 12,000-items peak (see Linacre, 2019).

### 4.1. Distribution of Blacklisted Journals as Per Crawford's Codes

Figure 1 shows the distribution of the 4,681 journals that we observed included in Cabell's blacklist as per Crawford's codes. As previously mentioned,

Crawford used codes in his Gray OA 2012-2016 dataset to tag the different types of questionable journals that he encountered in his examination of Beall's lists. These codes notably include: XU (unworkable), XX (unreachable), XO (Opaque; too difficult to count), XH (hybrid), and E (empty). This last code (Empty) is assigned to 3,469 out of the 4,681 journals that we observed included in Cabell's blacklist. This number represents 74.11% of our sampled journals included in the blacklist and 28.91 % of the total sum of journals in Cabell's blacklist, that is on the basis of the 12,000 journals peak it reached in early October 2019 (see Linacre, 2019).

After verifying whether Crawford's empty code still applied to these 3,469 journals, the number of empty journals that we observed included in the blacklist slightly decreased. This is what can be seen in Figure 2, which shows the distribution of these 3,469 journals as per the new codes they were assigned to – either one of Crawford's original codes which may be different from the one previously assigned by Crawford, or a new code: NE for "Not empty anymore". In this graph, the number of still truly empty journals amounts to 3,229, that is 26.91% of Cabell's blacklist as per its 12,000 journals peak. This proportion of empty journals is thus an absolute minimum since our sample does not examine the whole of Cabell's blacklist. The reason why some journals are not empty anymore does not specifically reveal a flaw in Crawford's tagging or methodology. Most of the journals that are not empty anymore (112) have published articles in the second half of 2016 or later, that is after Crawford performed his verifications. Moreover, a bit more than 100 journals previously tagged as empty by Crawford have now become unreachable (i.e. urls are not working anymore).

*Fig. 1: The distribution of journals that we observed included in Cabell's blacklist as per Crawford's codes.*

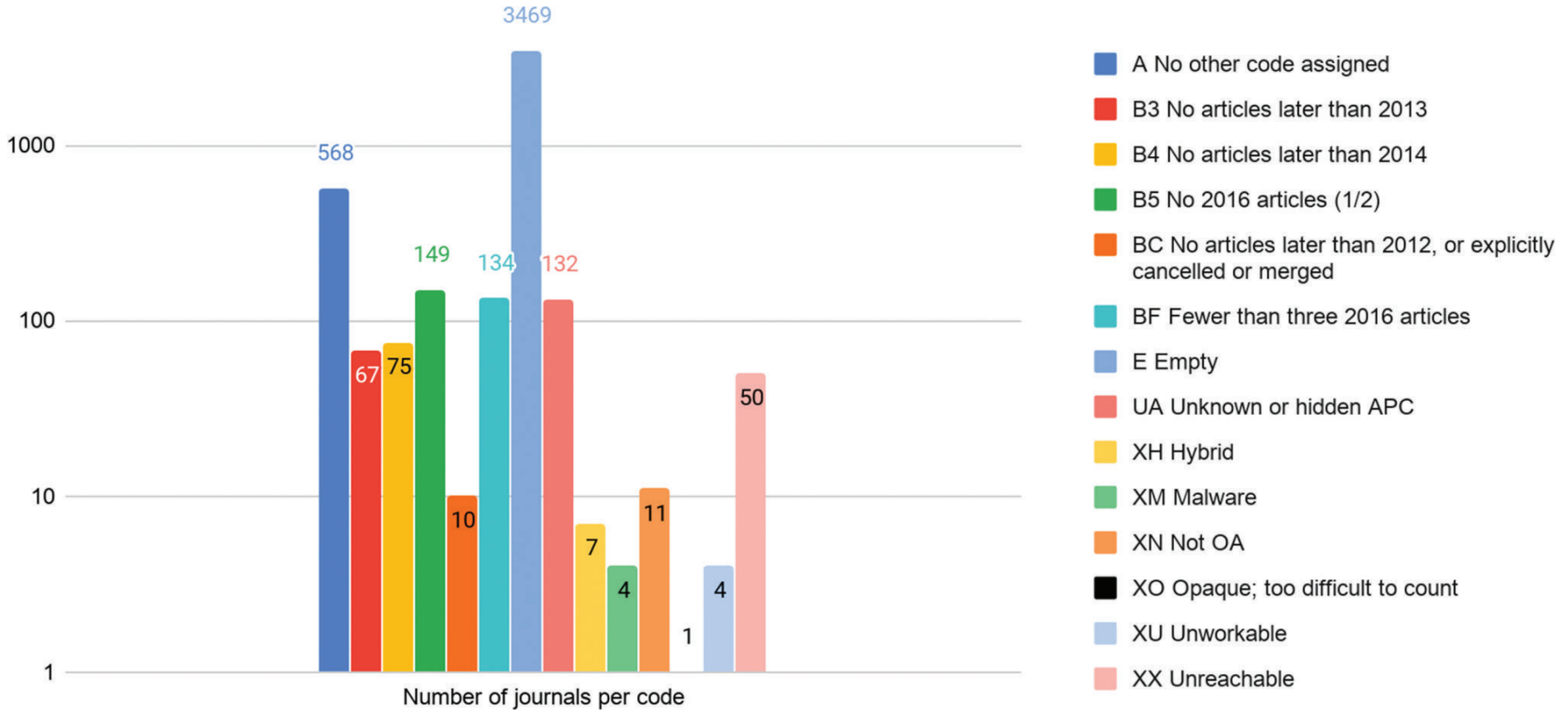

### 4.2. Distribution of Blacklisted Journals as Per Number of Violations

Figure 3 details the distribution of the number of journals that we observed included in the blacklist as per number of violations registered by Cabell. A bit more than half of these journals (51.31%) are included in the blacklist on the basis of 7 violations. The extreme segments of Cabell's violation spectrum in our sample, i.e. 1 to 3 violations and 11 to 15 violations, account for, respectively, 1.41% (66) and 2.2% (103) of the total number of journals that we observed included in the blacklist (4,681).

While proceeding to cross-verifications for journals blacklisted on the basis of a high number of violations, we were surprised to see repeated statements of violations, that is identical violations were recorded multiple times (see example in Figure 4). We therefore checked the frequency of this phenomenon in a journals segment that we virtually created on a high number of violations (11 to 15 violations). After checking the 103 journals included in the blacklist on the basis of 11 to 15 violations, we observed 11 instances of journals indexed with repeated statements of violations (see Figure 5). This represents 10.68% of the the said journals segment. Figure 5 shows the list of the 11 journals that we identified as having identical violations recorded multiple times and the corrected number of violations for each journal, that is, minus identical violations. Details of those identical violations for each journal can be found in the 'Detail' column of our dataset.[8] As a result of this irregularity, the data in Figure 3 is slightly inaccurate.

*Fig. 2: New code distribution of the 3,469 journals originally tagged as empty by Crawford that we observed included in Cabell's blacklist.*

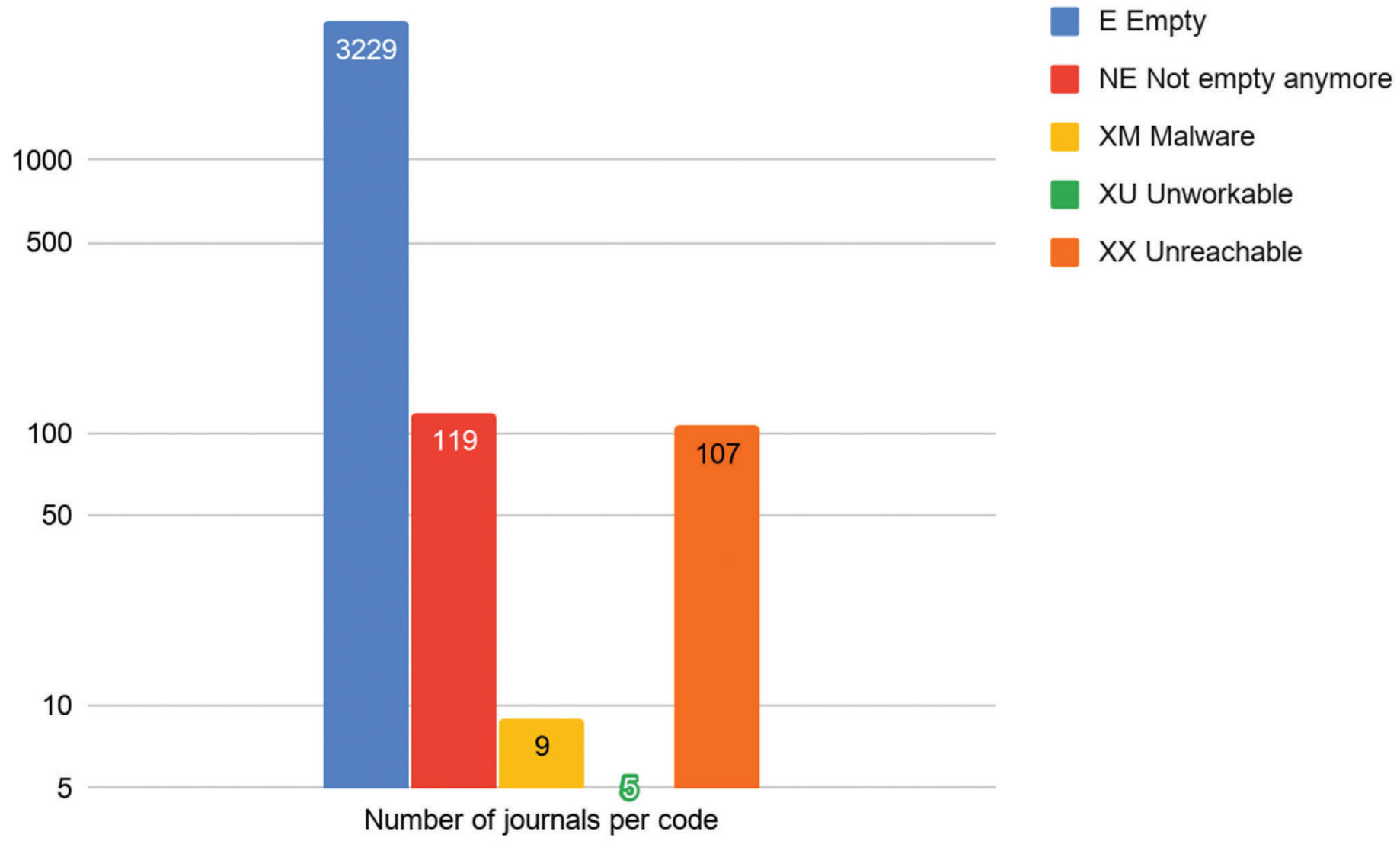

*Fig. 3: Number of journals included in Cabell's blacklist as per number of violations (as per Cabell's data).*

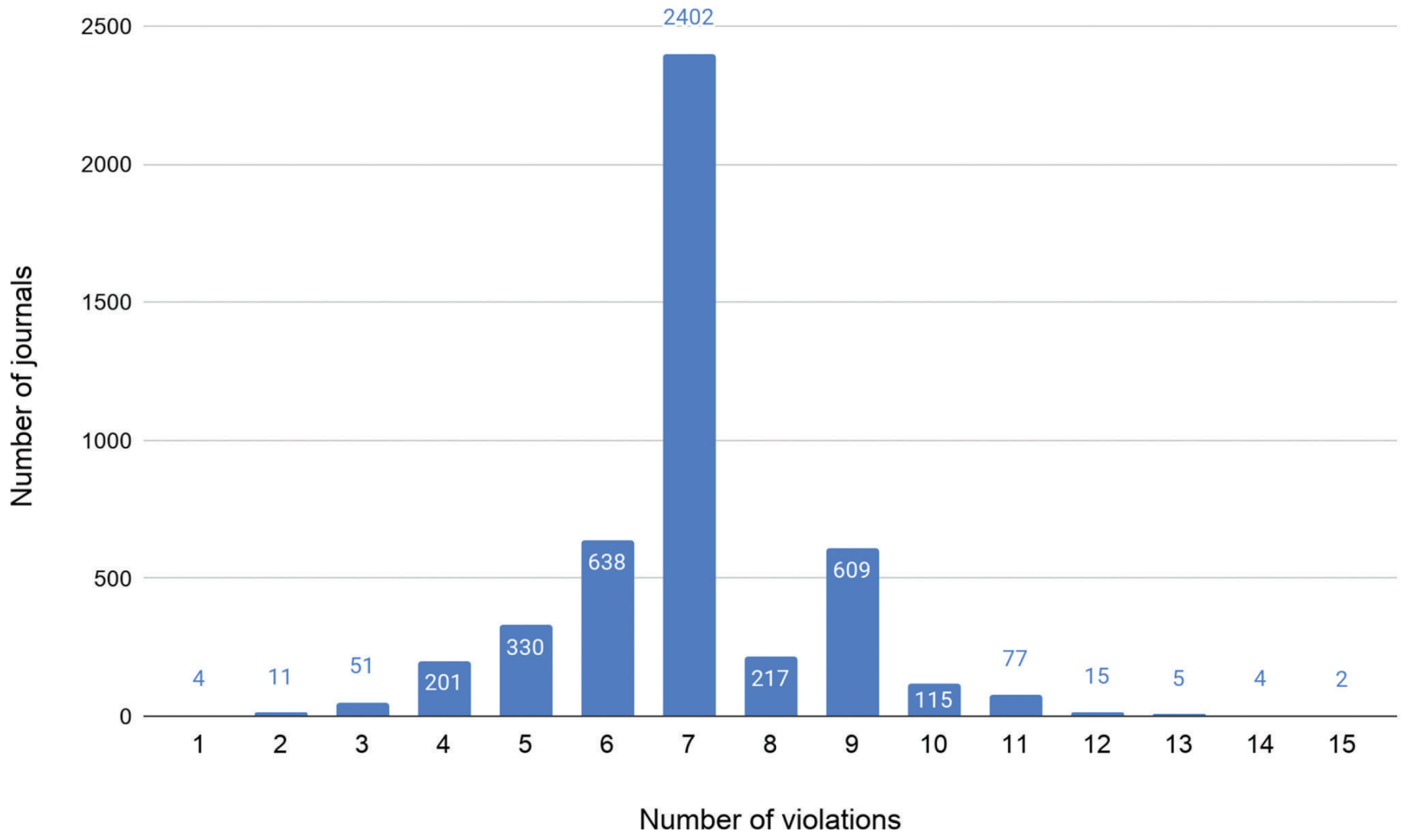

### 4.3. Detail and Frequency of Violation Combinations for Journals Blacklisted on the Basis of 1 to 3 Criteria

The segment of journals that we observed in the blacklist on the basis of 1 to 3 criteria contains 66 journals, for which 29 different types of violation combinations have been used. Table 2 below details these combinations and the number of journals blacklisted for each combination, for which details of individual violations are provided. For purposes of clarity, each combination was assigned a code (a to ac). The most frequent combinations within this segment of 66 journals correspond to the following codes: g (9 journals; 3 violations – 1 severe, 2 moderate), n (5 journals; 3 violations – 1 minor, 2 moderate), o (5 journals; 3 violations – 1 severe, 2 moderate), b (4 journals; 2 violations – 1 severe, 1 minor), and a (4 journals; 1 violation – severe).

*Fig. 4: An example of a journal's detailed entry in Cabell's blacklist with identical violations recorded multiple times. Screenshot captured on September 8th, 2019.*

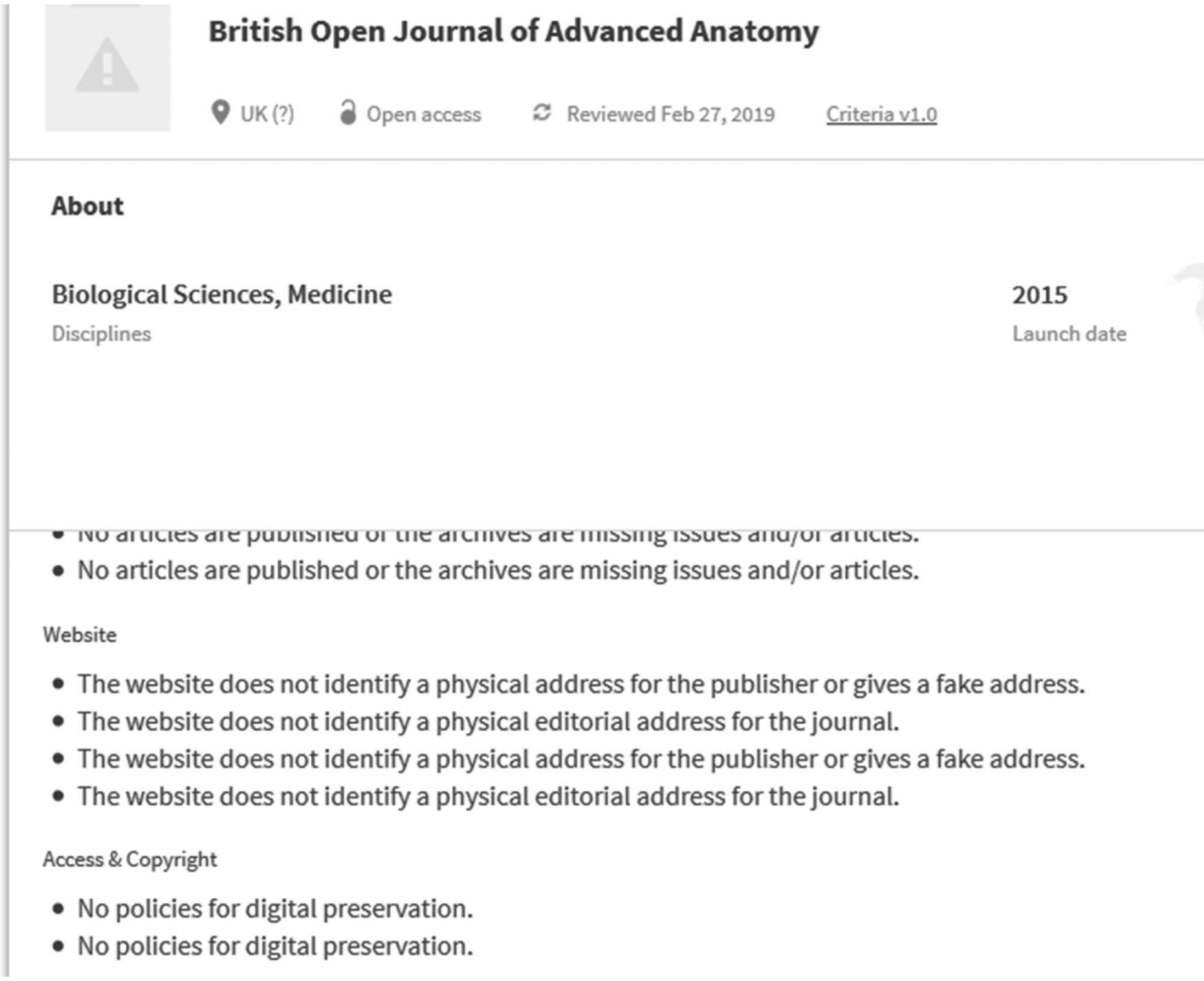

## 5. Discussion

### 5.1. Empty Journals

The very high number of empty journals indexed in Cabell's blacklist (3,229) raises serious questions about the ways in which they prioritise journals for inclusion and their willingness to provide an up-to-date and useful blacklist to the scholarly community. By definition, empty fraudulent journals have not succeeded in harming researchers or the scholarly community since their preying has not been effective yet. It is possible to argue that some of these empty journals may have already managed to collect APCs from inattentive or desperate authors without publishing their articles. But for two obvious reasons, we find this possibility to be very unlikely and most probably insignificant statistically speaking. First, authors will most likely not consider an empty journal to submit an article because it will not help them increase their visibility or prestige. Second, a fake journal must know that they won't attract many submissions if their publishing outlet remains empty. It may also be argued that some of these journals may well turn into effective fraudulent journals some day by managing to get articles published. But this seems unlikely since the 3,229 empty journals that we observed as still empty have been tagged as such since at least December 1st, 2016, that is when Crawford posted his Gray OA 2012-2016 dataset on the Figshare repository.

*Fig. 5: Journals that we observed containing identical violations recorded multiple times, with their corrected number of violations.*

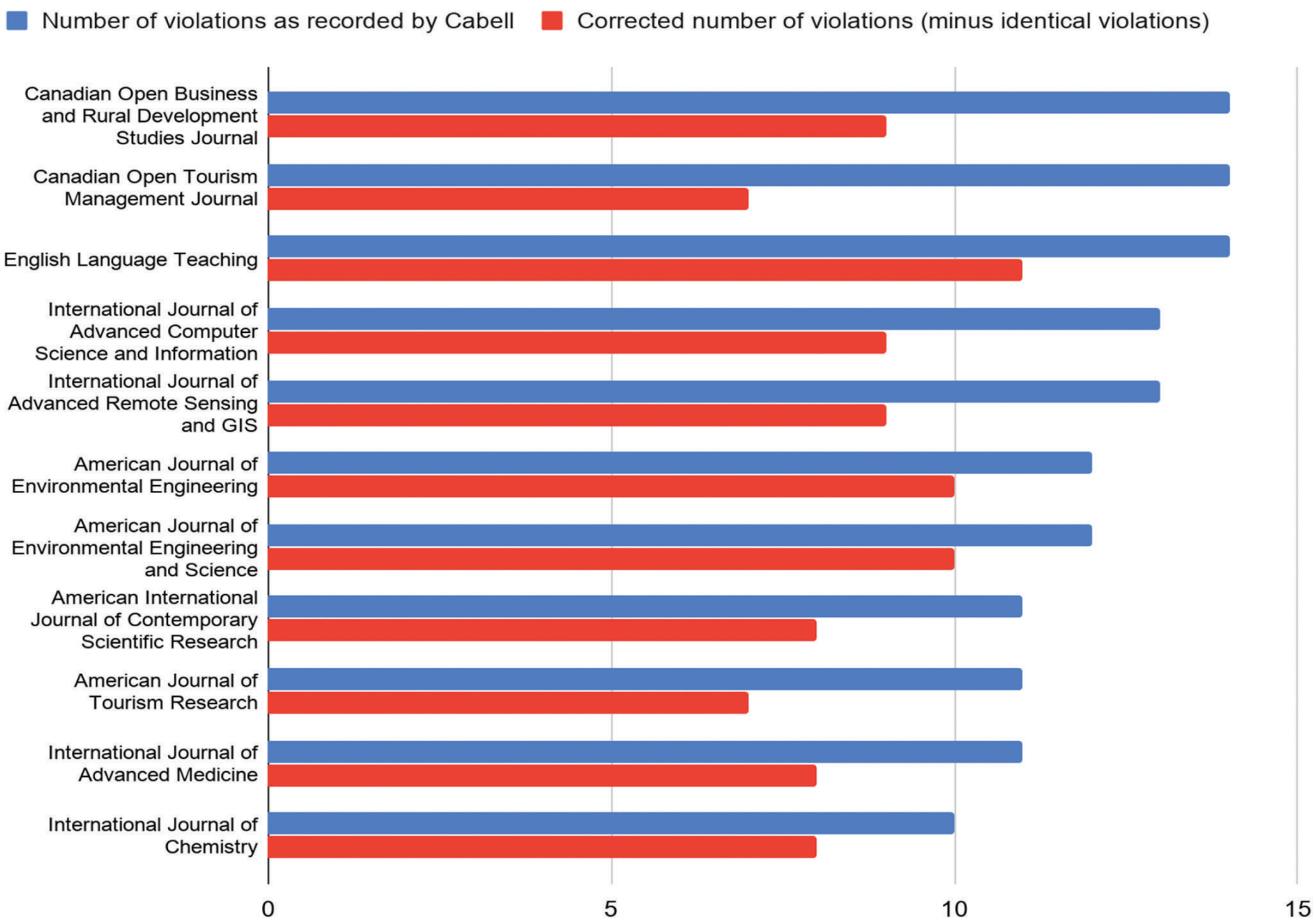

*Table 2: Number and detail of violation combinations for journals we observed included in Cabell's blacklist on the basis of 1 to 3 criteria.*

| Code | Number of violations used | Violation(s) detail and degree of severity | Number of journals blacklisted according to detail |
|---|---|---|---|
| **a** | 1 | No articles are published or the archives are missing issues and/or articles. (severe) | 4 |
| **b** | 2 | The journal states there is an APC or another fee but does not give information on the amount or gives conflicting information. (severe) | 4 |
| | | Poor grammar and/or spelling on the journal or publisher's website. (minor) | |
| **c** | 2 | No articles are published or the archives are missing issues and/or articles (severe) | 3 |
| | | No policies for digital preservation. (moderate) | |
| **d** | 2 | Falsely claims indexing in well-known databases (especially SCOPUS, DOAJ, JCR, and Cabells). (severe) | 2 |
| | | The publisher displays prominent statements that promise rapid publication and/or unusually quick peer review (less than 4 weeks). (moderate) | |
| **e** | 2 | No editor or editorial board listed on the journal's website at all. (severe) | 1 |
| | | No way to access articles (no information on open access or how to subscribe). (moderate) | |
| **f** | 2 | The same article appears in more than one journal. (severe) | 1 |
| | | No policies for digital preservation. (moderate) | |
| **g** | 3 | No editor or editorial board listed on the journal's website at all. (severe) | 9 |
| | | The publisher displays prominent statements that promise rapid publication and/or unusually quick peer review (less than 4 weeks). (moderate) | |
| | | No policies for digital preservation. (moderate) | |

*Table 2 (continued)*

| Code | Number of violations used | Violation(s) detail and degree of severity | Number of journals blacklisted according to detail |
|---|---|---|---|
| **h** | 3 | No editor or editorial board listed on the journal's website at all. (severe) | 3 |
| | | No articles are published or the archives are missing issues and/or articles. (severe) | |
| | | The publisher displays prominent statements that promise rapid publication and/or unusually quick peer review (less than 4 weeks). (moderate) | |
| **i** | 3 | No editor or editorial board listed on the journal's website at all. (severe) | 2 |
| | | No articles are published or the archives are missing issues and/or articles. (severe) | |
| | | No policies for digital preservation. (moderate) | |
| **j** | 3 | The website does not identify a physical address for the publisher or gives a fake address. (minor) | 2 |
| | | The website does not identify a physical editorial address for the journal. (minor) | |
| | | No articles are published or the archives are missing issues and/or articles. (severe) | |
| **k** | 3 | The website does not identify a physical editorial address for the journal. (minor) | 2 |
| | | The publisher displays prominent statements that promise rapid publication and/or unusually quick peer review (less than 4 weeks). (moderate) | |
| | | Authors are published several times in the same journal and/or issue. (moderate) | |
| **l** | 3 | No articles are published or the archives are missing issues and/or articles.(severe) | 1 |
| | | The publisher displays prominent statements that promise rapid publication and/or unusually quick peer review (less than 4 weeks). (moderate) | |
| | | No policies for digital preservation. (moderate) | |

*Table 2 (continued)*

| Code | Number of violations used | Violation(s) detail and degree of severity | Number of journals blacklisted according to detail |
|---|---|---|---|
| **m** | 3 | No articles are published or the archives are missing issues and/or articles. (severe) | 1 |
| | | The journal's website does not have a clearly stated peer review policy. (moderate) | |
| | | No policies for digital preservation. (moderate) | |
| **n** | 3 | The publisher hides or obscures relationships with for-profit partner companies. (moderate) | 5 |
| | | The journal or publisher uses a virtual office or other proxy business as its physical address. (minor) | |
| | | The publisher displays prominent statements that promise rapid publication and/or unusually quick peer review (less than 4 weeks). (moderate) | |
| **o** | 3 | Falsely claims indexing in well-known databases (especially SCOPUS, DOAJ, JCR, and Cabells). (severe) | 5 |
| | | The publisher displays prominent statements that promise rapid publication and/or unusually quick peer review (less than 4 weeks). (moderate) | |
| | | Authors are published several times in the same journal and/or issue. (moderate) | |
| **p** | 3 | The owner/Editor of the journal or publisher falsely claims academic positions or qualifications. (severe) | 2 |
| | | The journal or publisher uses a virtual office or other proxy business as its physical address. (minor) | |
| | | No policies for digital preservation. (moderate) | |

*Table 2 (continued)*

| Code | Number of violations used | Violation(s) detail and degree of severity | Number of journals blacklisted according to detail |
|---|---|---|---|
| **q** | 3 | No editor or editorial board listed on the journal's website at all. (severe) | 2 |
| | | Falsely claims indexing in well-known databases (especially SCOPUS, DOAJ, JCR, and Cabells). (severe) | |
| | | The publisher displays prominent statements that promise rapid publication and/or unusually quick peer review (less than 4 weeks). (moderate) | |
| **r** | 3 | The publisher hides or obscures relationships with for-profit partner companies. (moderate) | 2 |
| | | No editor or editorial board listed on the journal's website at all. (severe) | |
| | | The publisher displays prominent statements that promise rapid publication and/or unusually quick peer review (less than 4 weeks). (moderate) | |
| **s** | 3 | Falsely claims indexing in well-known databases (especially SCOPUS, DOAJ, JCR, and Cabells). (severe) | 2 |
| | | The publisher displays prominent statements that promise rapid publication and/or unusually quick peer review (less than 4 weeks). (moderate) | |
| | | The same article appears in more than one journal. (severe) | |
| **t** | 3 | Falsely claims indexing in well-known databases (especially SCOPUS, DOAJ, JCR, and Cabells). (severe) | 2 |
| | | The journal states there is an APC or another fee but does not give information on the amount or gives conflicting information. (severe) | |
| | | No policies for digital preservation. (moderate) | |

*Table 2 (continued)*

| Code | Number of violations used | Violation(s) detail and degree of severity | Number of journals blacklisted according to detail |
|---|---|---|---|
| **u** | 3 | No articles are published or the archives are missing issues and/or articles. (severe)<br><br>The journal or publisher uses a virtual office or other proxy business as its physical address. (minor)<br><br>No policies for digital preservation. (moderate) | 2 |
| **v** | 3 | No policies for digital preservation. (moderate)<br><br>No articles are published or the archives are missing issues and/or articles. (severe)<br><br>Falsely claims indexing in well-known databases (especially SCOPUS, DOAJ, JCR, and Cabells). (severe) | 2 |
| **w** | 3 | Information received from the journal does not match the journal's website. (severe)<br><br>The website does not identify a physical address for the publisher or gives a fake address. (minor)<br><br>The publisher or its journals are not listed in standard periodical directories or are not widely catalogued in library databases. (minor) | 1 |
| **x** | 3 | The journal includes scholars on an editorial board without their knowledge or permission. (severe)<br><br>The journal states there is an APC or another fee but does not give information on the amount or gives conflicting information. (severe)<br><br>Poor grammar and/or spelling on the journal or publisher's website. (minor) | 1 |
| **y** | 3 | The journal uses misleading metrics (i.e., metrics with the words "impact factor" that are not the Thomson Reuters Impact Factor). (severe)<br><br>The journal's website does not have a clearly stated peer review policy. (moderate)<br><br>No policies for digital preservation. (moderate) | 1 |

*Table 2 (continued)*

| Code | Number of violations used | Violation(s) detail and degree of severity | Number of journals blacklisted according to detail |
|---|---|---|---|
| **z** | 3 | No articles are published or the archives are missing issues and/or articles. (severe) | 1 |
| | | The journal states there is an APC or another fee but does not give information on the amount or gives conflicting information. (severe) | |
| | | Poor grammar and/or spelling on the journal or publisher's website. (minor) | |
| **aa** | 3 | No editor or editorial board listed on the journal's website at all. (severe) | 1 |
| | | The publisher displays prominent statements that promise rapid publication and/or unusually quick peer review (less than 4 weeks). (moderate) | |
| | | Authors are published several times in the same journal and/or issue. (moderate) | |
| **ab** | 3 | The owner/Editor of the journal or publisher falsely claims academic positions or qualifications. (severe) | 1 |
| | | The publisher displays prominent statements that promise rapid publication and/or unusually quick peer review (less than 4 weeks). (moderate) | |
| | | Authors are published several times in the same journal and/or issue. (moderate) | |
| **ac** | 3 | The publisher hides or obscures relationships with for-profit partner companies. (moderate) | 1 |
| | | The journal's website does not have a clearly stated peer review policy. (moderate) | |
| | | The publisher displays prominent statements that promise rapid publication and/or unusually quick peer review (less than 4 weeks). (moderate) | |

We do not wish to suggest that these empty journals should not be indexed, but maybe they deserve a category of their own so as to not artificially boost the size of Cabell's blacklist, which is of course used as a powerful sales argument by the company (Linacre, 2019). Cabell's blacklist would be much more useful to stakeholders in scholarly communications if it prioritised truly fraudulent journals that are not empty for inclusion, that is journals which have already and effectively harmed the scholarly community and continue to do so. Unfortunately, this does not seem to be the case at all, as Cabell seems to continue to massively review and index empty journals. Our data indeed indicates that out of the 822 journals that Cabell reviewed in 2019, 687 (83.6%) are empty journals (after verification of Crawford's codes).

### 5.2. Weighing Method, Problematic Violations Combinations, and Minimal Requirements for Inclusion

The high number of empty journals included in Cabell's blacklist is most likely due to their use of the violation "No articles are published or the archives are missing issues and/or articles" (severe) for reviewing. This violation appears in both of Cabell's criteria versions (v1.0 and v1.1) and is considered 'severe' in their v1.1 criteria. Quite surprisingly, our data indicates that this is also the only violation that is used on its own to blacklist journals (see Table 2, code a). According to our data, Cabell has blacklisted 4 journals on the basis of this unique violation. We seriously doubt that this sole criterion may, on its own, count for over the 100 points that Cabell has defined as its threshold score for inclusion in the blacklist (Bisaccio, 2018). Moreover, if this sole criterion can be used to blacklist journals, one might wonder why it has not been applied to the 3,229 previously mentioned truly empty journals included in Cabell's blacklist, all of which are recorded in the blacklist on the basis of several violations (ranging from 2 to 14). In fact, this surprising element indicates that Cabell's reviewing method may vary from title to title. Moreover, this raises questions regarding cross-verifications in Cabell's workflow.

We also find the above violation unclear and problematic, and believe it should not be used on its own to mark journals as fraudulent. If no articles are published in a journal, then said journal's potential 'predatory' character is not real or achieved yet.[9] Moreover, the idea of missing issues entailed in this violation may also be problematic if a journal does not explicitly state its publishing schedule or adopts a continuous publication model, which may render this aspect of "missing issues" not applicable to all journals. Finally, the idea of missing articles entailed in this violation could also be clarified. To the best of our knowledge, very few journals explicitly state the number of expected articles to be published in a single issue. We assume that what Cabell means by this is that journals which repeatedly publish issues with only one or two articles can be considered dubious. For reasons of clarity, it probably would have been better for Cabell to break down this violation into several criteria and to provide a separate category for truly empty journals.

Another problematic criterion used by Cabell is the following "no policies for digital preservation" (moderate). This criterion was used by Jeffrey Beall for identifying fraudulent journals before it was used by Cabell. And its relevance has already been seriously questioned in a detailed study surveying the subjectivity of Beall's criteria (Olivarez et al., 2018). Olivarez et al. show that 22 out of 87 authentic and legitimate journals fail Beall's test on the basis of this only criterion, which constitutes "the most failed criterion" of their study (Olivarez et al., 2018). Moreover, this study reveals a rather high percentage of disagreement regarding the criterion about digital preservation. In fact, the analysts who were asked to review a set of journals according to Beall's criteria only agreed 60% of the time in regard to to this specific criterion. To be fair, this article had not yet been published when Cabell launched its blacklist in early 2017, but this information was available when Cabell started working on their v 1.1 version, which officially went into effect on May 13th, 2019 (Toutloff, 2019b).

This criterion does not reflect the very diverse nature underlying the ecosystem of academic journal publishing. Many authentic and legitimate journals simply do not have explicit policies regarding digital preservation, especially small publishing venues that have been created and are maintained by associations, departments, or libraries, sometimes at great effort and cost. This does not mean that the people running these journals do not care about digital preservation, though. Rather, they usually consider that journal content uploaded on an institutional or scientific society server already guarantees digital preservation. This is why some libraries or societies maintaining these journals do not always deem it necessary to explicitly state or develop a digital preservation policy, let alone pay for third-party services that provide such digital preservation.

Despite the many existing reservations regarding this criterion ("no policies for digital preservation"), our data reveals that Cabell has used it in combination with the other previously discussed questionable criterion ("No articles are published or the archives are missing issues and/or articles") to blacklist 3 journals (see Table 2, code c). Of course, this represents a very small fraction of the total number of journals we observed included in the blacklist, but it raises further questions about Cabell's methodology and weighing process. Here again, it is very unlikely that the addition of points assigned to these two violations (1 moderate and 1 severe) may effectively reach Cabell's threshold of 100.

The above examples show that Cabell's minimal requirements for inclusion in the blacklist are questionable and could certainly be improved. Other associations of questionable criteria could be pinpointed in this respect. For example, we find the combinations of violations comprised between the codes b-c and g-m to be problematic (see Table 2 for detail of violations combinations with degree of severity as assigned by Cabell for individual violations). Taken together, these combinations have been used to blacklist 27 journals. All of these combinations associate at least one of the two previously discussed problematic criteria – "No policies for digital preservation" (moderate) and "No articles are published or the archives are missing issues and/or articles" (severe) – with at least one of the following questionable criteria presented in Cabell's v1.1 (Toutloff, 2019b):

- "Poor grammar and/or spelling on the journal or publisher's website" (minor);
- "The publisher displays prominent statements that promise rapid publication and/or unusually quick peer review (less than 4 weeks)" (moderate);
- "No editor or editorial board listed on the journal's website at all" (severe);
- "The journal's website does not have a clearly stated peer review policy" (moderate).

We do not wish to suggest that all of these criteria are irrelevant, or that journals blacklisted on the basis of these questionable combinations of violations are not fraudulent. But, methodologically speaking, the ways in which these criteria are combined together make it hard to distinguish poor-quality journals from truly deceptive ones.

### 5.3. Risks of Unsystematic Comprehensive Reviewing

Other insights from Cabell's methodology can be deduced from looking at the bigger picture of the number of journals indexed in the blacklist as per number of violations (see Figure 3). This figure is very similar to Kyle Siler's graph presenting the distribution of journals according to the number of violations in the whole of Cabell's blacklist – as of May 2019, N = 11,450 (Siler, 2020, see Figure 1). Put into perspective with the total number of possible violations that Cabell uses to test journals – 64 criteria in its v1.0 (Toutloff, 2019a) and 74 criteria in its v1.1(Toutloff, 2019b) – this distribution of journals according to the number of violations raises questions as to whether Cabell's lists of criteria are always checked thoroughly for each reviewed title. In our sample, for example, more than half of the journals (51.3%) are blacklisted on the basis of 7 violations, that is, merely 10.9% and 9.46% of the total number of Cabell's violations, respectively in their v1.0 and v1.1 criteria versions.[10]

Performing exhaustive checks based on Cabell's criteria versions may be difficult for two reasons. First, some criteria listed by Cabell to identify fraudulent journals can be very hard and/or time-consuming to verify. A non-exhaustive list of these criteria included in Cabell's v1.1 (Toutloff, 2019b) would include:

- "Editorial board members (appointed over 2 years ago) have not heard from the journal at all since being appointed to the board" (moderate);
- "Inadequate peer review (i.e., a single reader reviews submissions; peer reviewers read papers outside their field of study; etc.)" (moderate);
- "The journal has been asked to quit sending emails and has not stopped" (moderate);
- "The number of articles published has increased by 75% or more in the last year" (moderate);
- "The number of articles published has increased by 50–74% in the last year" (moderate);
- "The number of articles published has increased by 25–49% in the last year" (minor).

Most of these criteria's verifiability largely depends on whether or not Cabell's reviewers are able to have 'insider knowledge' on a journal's procedures, which is most likely rare. Second, our sample shows a rather low ratio of criteria for inclusion in the blacklist when compared to the total number of criteria used in Cabell's v1.0 and v1.1. This might be the result of a tendency to privilege high-value violations to reach Cabell's threshold score more rapidly. Although this claim should be taken with caution, it might explain – at least in part – why Cabell has managed to rapidly increase the number of journals included in their blacklist. To lift this impression, it would be useful that Cabell be more transparent about its weighing process and reviewing methodology.

### 5.4. Irregular and Incomplete Data

We observed several irregularities during our collection of data and analysis. In addition to the previously mentioned recording of identical violations for 11 journal entries, these irregularities include: two multiple recordings of individual journal entries with a different number of violations, discrepancies between the reviewing dates registered by Cabell and the criteria version used for 18 journals, missing reviewing dates for 103 journals, and the use of outdated information regarding Cabell's misleading metrics criterion.

These irregularities, which we present and briefly discuss below, only affect a very small proportion of the journals that we observed included in the blacklist. Nevertheless, we thought it useful to detail them. After all, one of Cabell's objectives in building a blacklist is to expose fraudulent journals and publishers on the basis of their problematic or shaky procedures and/or policies. It is thus quite logical to adopt the same rigour regarding Cabell's own procedures and data collection methods, and to highlight their sometimes sloppy character.

The first irregularity that we observed is the repeated statements of violations at the level of individual journal entries, i.e. individual journal entries in Cabell's blacklist that display identical violations recorded multiple times (see Figure 4). In our sample, we observed this phenomenon for 11 journals (see Figure 5). But it is very likely that Cabell's blacklist contains many more such journals with repeated statements of violations. First, our study only covers 39% of the 12,000 journals included in Cabell's blacklist as of early October 2019. Second, the 11 journals that we observed contained repeated statements of violations account for 10.68 % of the number of journals included in the blacklist for which we checked the presence of identical violations (103 journals blacklisted on the basis of 11 to 15 violations). It is likely that the proportion of journals with identical violations is lower for the rest of Cabell's blacklist. Nevertheless, this raises further questions about the rigorous character of Cabell's reviewing procedure and weighing methods. It is not clear, for instance, whether repeated statements of violations impact a journal's score before it is included in the blacklist.

A second irregularity concerns identical journals recorded twice with a different number of violations, of which we found two instances (see Figure 6 and Figure 7). The journal *Biomedical Engineering Review*, published by KEI Journals, was first included on August 21, 2017 on the basis of 3 violations. It was later indexed a second time on February 8, 2018 on the basis of 7 violations (see Figure 6). The two reviews for this journal have 3 violations in common. Similarly, the journal *Advances in Materials Science and Applications*, published by World Academic Publishing, was first indexed on October 26, 2016 on the basis of 3 violations. It was later indexed a second time on June 26, 2017 on the basis of 4 violations (see Figure 7). The two reviews for this journal have only one violation in common. These discrepancies in number of violations recorded for identical journals suggest that Cabell's reviewing method and process is probably too subjective; it is very likely that these journals were diagnosed differently by different reviewers, or that they may have found some criteria to be vague or poorly operationalised.

*Fig. 6: Two different entries for identical journal Biomedical Engineering Review as they appear on Cabell's blacklist. Screenshot captured on September 8, 2019.*

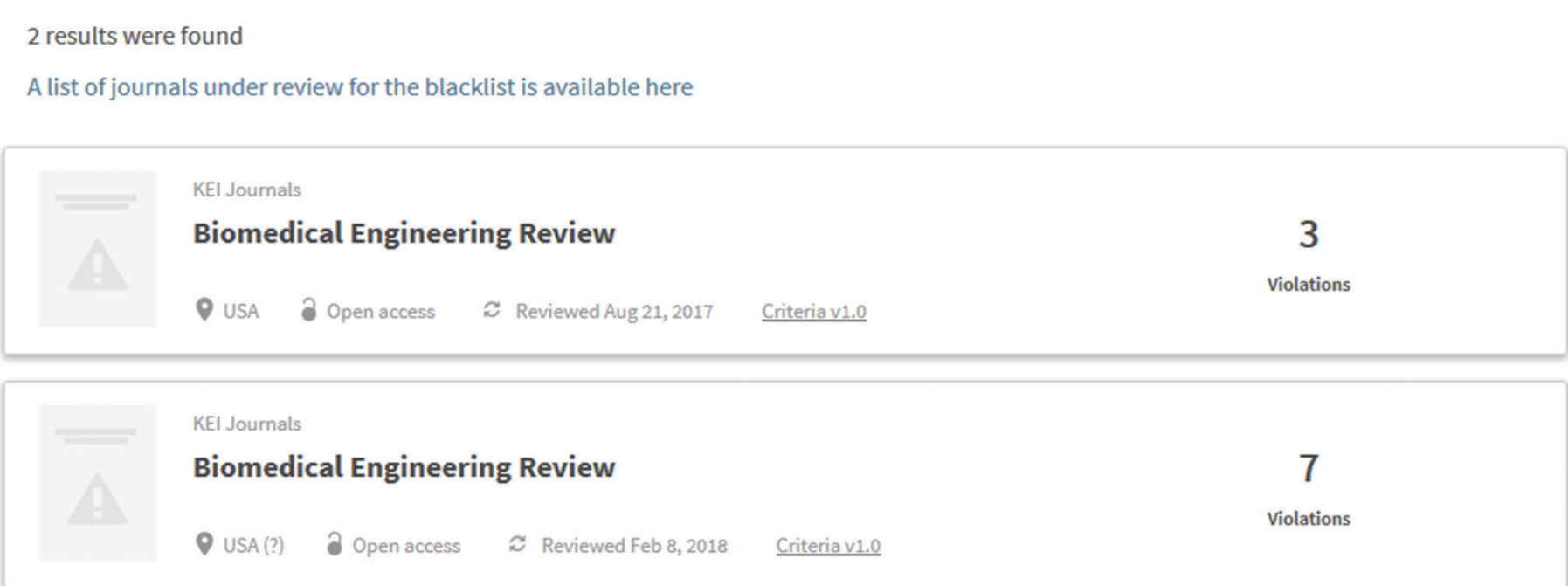

*Fig. 7: Two different entries for identical journal Advances in Materials Science and Applications as they appear on Cabell's blacklist. Screenshot captured on September 8, 2019.*

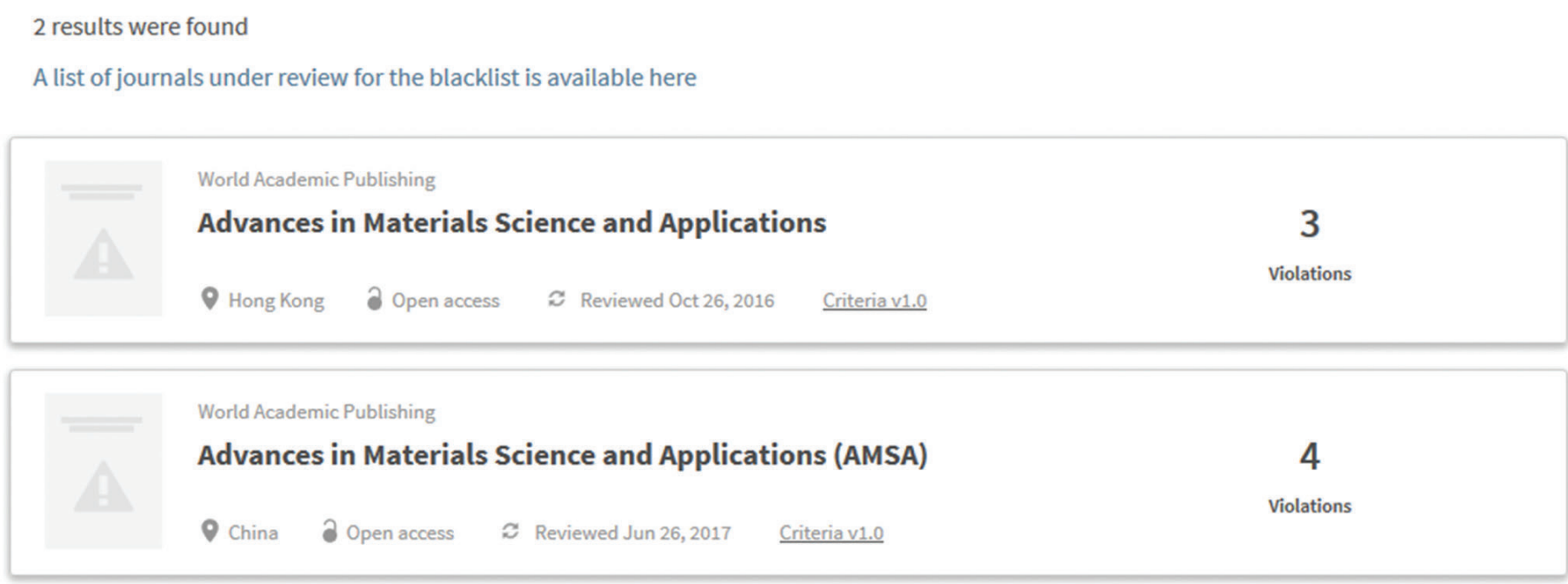

It is reasonable to believe that other instances of similar journals recorded multiple times exist in the rest of Cabell's blacklist for several reasons. First, we did not thoroughly check for the presence of such double entries in Cabell's blacklist. We identified these double entries when we collected the detail of violations journals included in the blacklist on the basis of 1 to 3 violations. Second, the indexing of identical journals by Cabell may be due to one severe limitation of their tool's search functions. At the time when this study was conducted, basic and advanced search functions did not support exact phrase searches. As a result, a standard journal title search may generate hundreds of results. This may have rendered the identification of a specific journal tricky or difficult for both users and, most likely, Cabell's reviewers themselves. Here again, the very possibility of such double entries points to questionable reviewing methods and procedures, including how journals are selected for review and Cabell's probable lack of cross-verifications before journals are effectively integrated in the blacklist.

Another irregularity concerns discrepancies regarding the date and/or criteria versions recorded by Cabell for several journals. Our sample shows that Cabell has failed to register a reviewing date for 103 journals.[11] Furthermore, our data indicates that 18 journals reviewed in 2016 or 2017 are included in the blacklist on the basis of Cabell's v1.1 criteria, (see Figure 8). This is quite puzzling since Cabell officially launched its v1.1 criteria version in March 2019 (Toutloff, 2019b). Obviously, this type of irregularity is not what one expects from a paywalled service that prides itself in providing the scholarly community with up-to-date and reliable data on fraudulent journals (Linacre, Bisaccio, & Earle, 2019).

*Fig. 8: An example of discrepancy between the reviewing date and the criteria version used for a journal entry as it appears on Cabell's blacklist. Screenshot captured on September 8th, 2019.*

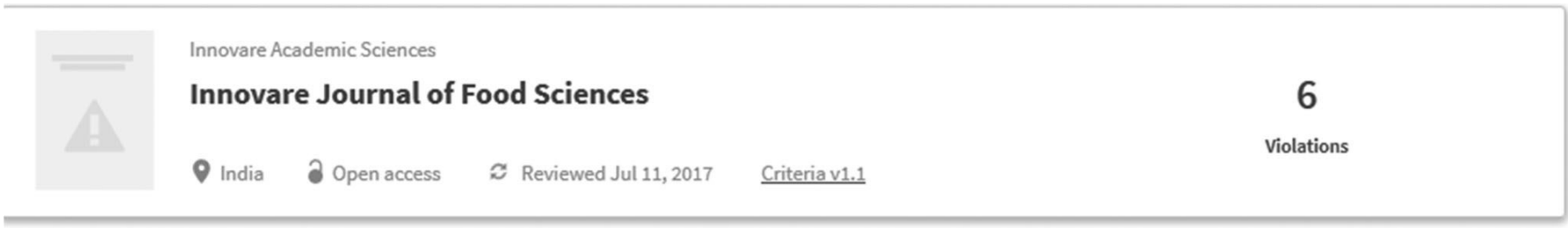

A final and rather surprising irregularity concerns Cabell's failure to update information regarding their misleading metrics criterion in their v1.1 criteria version, which went into effect in March 2019. The v.1.1 version of this criterion still refers to "the Thomson Reuters Impact Factor" (Toutloff, 2019b) – see lower screenshot in Figure 9. However, the company Clarivate Analytics has acquired the Thomson Reuters impact factor in 2017. Cabell's failure to update this information in their v.1.1 criteria version is even more surprising when one realises that they have updated this information for this equivalent criterion in their v1.0 criteria version (Toutloff, 2019a) – see upper screenshot in Figure 9.

*Fig. 9: Inconsistent updating of information relating to Cabell's misleading metrics criterion, which indicates a discrepancy between their v1.0 (top) and v1.1 (bottom) criteria versions. Screenshots captured on February 20, 2020.*

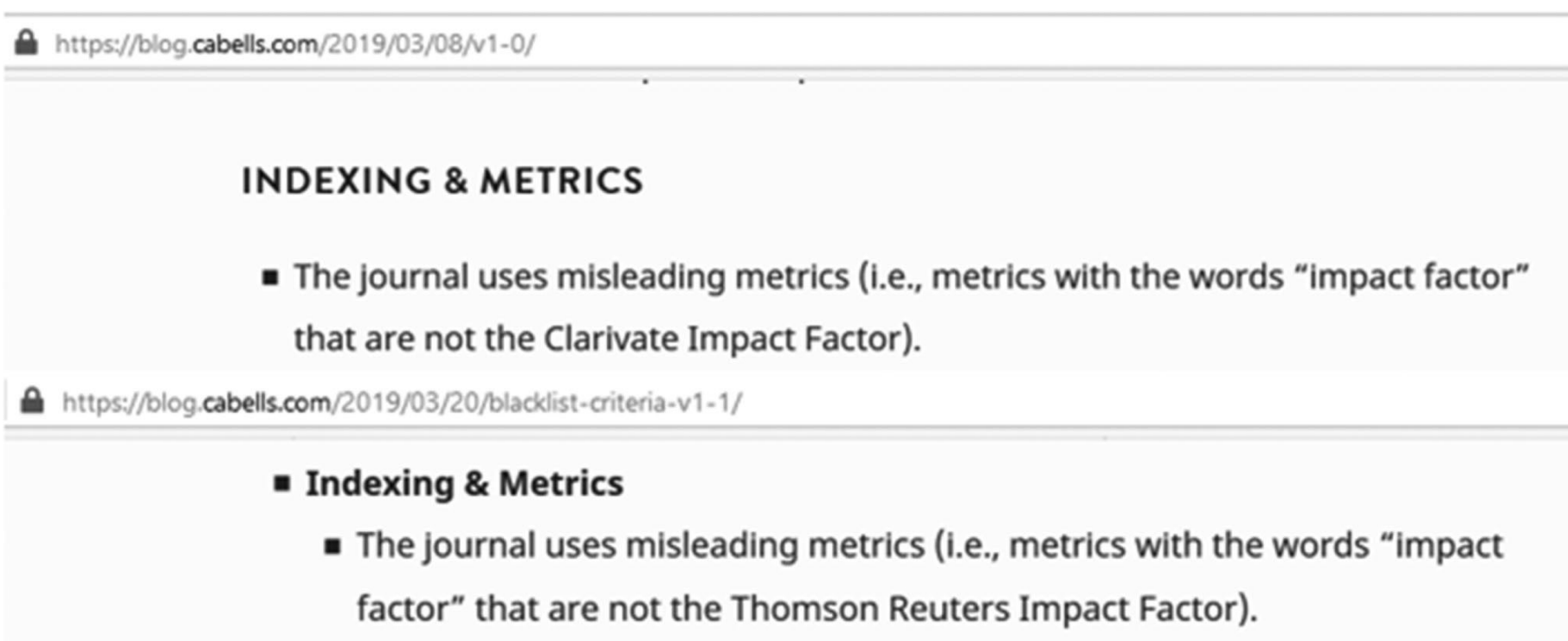

## 6. Conclusion and Perspectives

Cabell is the only available resource indexing fraudulent and deceptive journals by providing details of violations at the level of journal entry. To that extent, it is useful. Unfortunately, Cabell's methodology and weighing process for indexing journals and choosing which journals need to be reviewed can still very much be improved.

Based on our sample analysis, Cabell's significant indexing of empty journals (3,229), for example, shows that at least 26.9 % of its blacklist is made of empty shells, and therefore, somewhat empty promises for stakeholders in scholarly communications. We do not wish to suggest that Cabell should stop indexing such empty journals, but building a sub-list or a different category for these journals, and clearly alerting users of these journals' empty status might be a wise development. Moreover, it would be useful to prioritise journals that are not empty for inclusion in the blacklist.

Taking this issue of empty journals into consideration would also reflect a more nuanced understanding of predatory publishing article volumes. In a recent article, Cabell's executives remind us that "evidence suggests" that fraudulent and deceptive publishing practices "are growing in prevalence" (Linacre et al., 2019, p. 217) by citing the oft-quoted Shen and Björk paper (Shen & Björk, 2015), which reports a huge number of articles published in deceptive outlets for 2014. That "evidence", however, has been seriously challenged by Crawford (Crawford, 2017), precisely because more than half of the journals of his dataset are tagged as empty (52,9%).[12] And yet, nowhere in their article, do Linacre, Bissacio, and Earle mention Crawford's study or this issue of empty journals.

Overall, we find that Cabell's alleged added-value compared to other existing and freely available blacklists should currently be reconsidered. Cabell's many irregularities and approximations (double entries, recording of identical violations, erroneous recording of criteria versions, etc.) should be fixed so as to restore a trustworthy and a reliable image. To improve the tool, it would be necessary to propose a newer criteria version that is devoid of subjective criteria and of criteria that are very hard to objectively verify. Offering more transparency regarding weighing methods, the operationalisation of criteria, and how titles are selected for screening may also constitute welcome additions. Only then can Cabell's blacklist become a truly useful and reliable tool for stakeholders in scholarly communications.

Future scholarship may wish to further investigate Cabell's blacklist, its methodology, and procedures. Possible lines of inquiry include but are not limited to: examining the number of potential double entries and/or multiple recording of identical violations for individual journal entries in the whole of Cabell's blacklist, or in a random sample. Tracking the evolution of Cabell's indexing of empty journals and assessing their proportion in the blacklist may also be useful for stakeholders in scholarly communications, just like determining the proportion of criteria that are effectively used by Cabell to blacklist journals. Finally, evaluating how the results obtained in this study may have evolved after a certain period of time, and assessing to what extent Cabell may have fixed some of the issues that we identified could provide additional insightful data to assess how useful and reliable Cabell's blacklist is.

## Data

Our dataset is available at the LIBER Quarterly Dataverse (part of the Harvard Dataverse), at https://doi.org/10.7910/DVN/IWJIRN. It can also be found at https://doi.org/10.5281/zenodo.4007501.

## Notes

[1] Although the use of the term 'predatory' is widely used in scholarship, it has repeatedly been called into question (see Anderson, 2015; Eriksson & Helgesson, 2018). We have decided to privilege adjectives such as fraudulent, fake, or deceptive rather than predatory for two reasons. First, some legitimate commercial publishers can very much be considered to "prey" on researchers and libraries as they request high APCs or subscription fees (see Rentier, 2018, p.25), or both. Second, some authors publishing in fake journals "are aware that the journals do not adhere to accepted standards but choose to publish in them anyway, hence they are not 'prey'" (Laine & Winker, 2017); they deliberately choose to publish in such fake journals for reasons that include fierce competition and a desire of career advancement (Shaghaei et al., 2018).

[2] We are aware that using the terms blacklist and whitelist may have some racist overtones for some readers (Houghton & Houghton, 2018). We would like to stress that our use of these terms is not meant to perpetuate racist culture. Rather, we use them because they are pervasive in the literature dealing with deceptive and fake academic publishing outlets. Moreover, our object of study in this paper uses this appellation (Cabell's Blacklist).

[3] Ulrich's periodical directory covers more than 400,000 serial publications and is the source of bibliographic and provider information in Ulrichsweb™.

[4] The last version of Beall's blog as harvested by the Internet Archive on January 3, 2017, is available at: http://web.archive.org/web/20170103170903/https://scholarlyoa.com/.

[5] Crawford's article count for 2016 takes only the January-June period into consideration.

[6] Detail of codes can be found in the "Codes" tab appearing at the bottom of Crawford's dataset (2016), which is available at: https://doi.org/10.6084/m9.figshare.4275860.v1.

[7] We would like to thank these primary data collectors: Olivier Borsus, Valérie Danthine, Ariane Ghislain, Thierry Jacques, Fabienne Prosmans, Sandra Rizzo, and Simona Stirbu.

[8] To do so, you can apply the filter 'yes' in the column "Identical violations listed multiple times". Alternatively, you can apply a filter in the journal column in inserting choosing one or several names of the journals mentioned in Figure 5.

[9] The "no articles published" part of this violation also raises questions concerning Cabell's possible reviewing of new authentic journals which may not have published any material yet. We assume that Cabell does not review newly launched journals but, for reasons of clarity, this may also be made explicit.

[10] During our two-week trial period, we were not able to record details of violations for a significant number of journals. As a result, we were not able to determine whether or not some of Cabell's criteria may never be used to blacklist journals.

[11] Detail of journals for which Cabell has failed to register a reviewing date can be found in our dataset using the filter "no date mentioned" in the "Last review date" column.

[12] Crawford (Crawford, 2017) argues that Shen and Björk's numbers are grossly overestimated. While we agree with this, it should also be said that Crawford's own numbers are probably underestimated considering that, as our study shows, a number of journals that he tagged as empty, and therefore did not take into consideration in his estimations, are not empty anymore.